\documentclass[aps,twocolumn,showpacs,floatfix]{revtex4}

\usepackage{graphicx}
\usepackage{epsf}
\usepackage{amsmath}

\newcommand{\ul}{\underline}
\newcommand{\kB}{k_{\mathrm{B}}}

\begin{document}
\date{\today}

\title{Origin of asymmetric reversal modes in
  ferromagnetic/antiferromagnetic multilayers}

\author{B.\ Beckmann, U.\ Nowak, and K.\ D.\ Usadel}
\affiliation{Theoretische Tieftemperaturphysik,
  Universit\"{a}t Duisburg-Essen, 47048 Duisburg, Germany}

\begin{abstract}
  Experimentally an asymmetry of the reversal modes has been found in
  certain exchange bias systems.  From a numerical investigation of
  the domain state model evidence is gained that this effect depends
  on the angle between the easy axis of the antiferromagnet and the
  applied magnetic field.  Depending on this angle the ferromagnet
  reverses either symmetrically, e.\ g.\ by a coherent rotation on
  both sides of the loop, or the reversal is asymmetric with a non
  uniform reversal mode for the ascending branch, which may even yield
  a zero perpendicular magnetization.
\end{abstract}

\pacs{75.70.Cn, 75.40.Mg, 75.50.Lk, 85.70-w}

\maketitle

For compound materials consisting of a ferromagnet (FM) in contact
with an antiferromagnet (AFM) a shift of the hysteresis loop along the
magnetic field axis can occur which is called exchange bias (EB).
Often, this shift is observed after cooling the entire system in an
external magnetic field below the N\'eel temperature $T_{\mathrm{N}}$
of the AFM (for a review see \cite{noguesJMMM99}). The role of the AFM
is to provide at the interface a net magnetization which is stable
during reversal of the FM consequently shifting the hysteresis
loop. The key for understanding EB is to understand this stability. 

In the approach of Malozemoff \cite{malozemoffPRB87} due to interface
roughness domain walls in the AFM perpendicular to the FM/AFM
interface are supposed to occur during cooling in the presence of the
magnetized FM resulting in a small net magnetization at the FM/AFM
interface. However, the formation of such walls and the stability of
the interface magnetization has never been proven. Additionally, the
formation of domain walls exclusively due to interface roughness is
energetically unfavorable and therefore unlikely to occur. Therefore,
other approaches have been developed where a domain wall forms in the
AFM parallel to the interface while the magnetization of the FM
rotates \cite{mauriJAP87,KoonPRL97}.  However, it was shown by
Schulthess and Butler \cite{schulthessPRL98} that in this model EB
vanishes if the motion of the spins in the AFM is not restricted to a
plane parallel to the film as was done in Koon's work.  To obtain EB
Schulthess and Butler assumed uncompensated AFM spins at the
interface. But their occurrence and stability during a hysteresis loop
is not {\em explained}, neither in their model nor in other similar
models \cite{stilesPRB99,kiwiEL97}.

In a recent experiment Milt\'enyi et al.\ \cite{miltenyiPRL00} showed
that it is possible to strongly influence EB in Co/CoO bilayers by
diluting the AFM CoO layer, i.\ e.\ by inserting
non-magnetic substitutions (Co$_{1-x}$Mg$_x$O)
or defects (Co\( _{1-y}\)O) 
not at the FM/AFM interface, but rather throughout the volume
part of the AFM.  In the same letter it was shown that a corresponding
theoretical model, the domain state (DS) model, investigated by Monte Carlo
simulations shows a behavior very similar to the experimental results.
According to these findings the observed EB has its origin in a DS in
the AFM which occurs during cooling and which is stabilized 
by non-magnetic defects  carrying a net magnetization at the AFM interface.
Later it was shown that a variety of experimental facts associated
with EB can be explained within this DS model
\cite{nowakJMMM02,nowakPRB02,kellerPRB02}.  The importance of defects
for the EB effect is also confirmed by recent experiments on
Fe$_x$Zn$_{1-x}$F$_2$/Co bilayers \cite{shiJAP02} and by
investigations \cite{mewesAPL00,mouginPRB01,misraJAP03} where it was
shown that it is possible to modify EB by means of irradiating an
FeNi/FeMn system by He ions in presence of a magnetic field. 
On the
other hand, models which assume the formation of a domain wall
parallel to the interface during field cooling cannot explain
these findings.  A domain wall parallel to the interface stores the
energy for redirecting the FM magnetization like a spring that is
wound up and non-magnetic defects will rather suppress this spring
effect leading to a decrease of EB with increasing defect
concentration. Further support for the relevance of domains in EB
systems is given by a direct spectroscopic observation of AFM domains
\cite{noltingNATURE00,ohldagPRL01}.

An important outstanding problem in EB systems is an explanation of
the asymmetry of the reversal mode during hysteresis observed in some
EB systems, either directly as an asymmetry of the shape of the
hysteresis loop or -- more detailed -- by neutron scattering
techniques \cite{fitzsimmonsPRL00,raduJMMM02}.  Even though in recent
simulations an asymmetry has been observed \cite{suessPRB03}, a
systematic theoretical investigation of this effect is still lacking.
In the following we will show that the occurrence of an asymmetry
depends on the angle between magnetic field and easy axis of the AFM.
A systematical variation of this angle reveals a rich variety of
different reversal modes not seen so far in simulations or in
theoretical investigations. These findings are not only important for
a deeper understanding of the EB mechanism but also essential for a
correct interpretation of the complex reversal behavior observed
experimentally.

We consider the so-called DS model which was introduced recently, for
a detailed discussion of its properties see \cite{nowakPRB02}. In the
following we have one FM monolayer exchange coupled to a diluted AFM
film consisting of 3 monolayers with dilution $p_{int}=0.5$ for the
interface layer and $p_{vol}=0.6$ for the remaining two layers (The
geometry is sketched in \cite{nowakPRB02}).  We deliberately chose
these values since these yielded the largest EB in previous
simulations.

The FM is described by a classical Heisenberg model (spin variable
$\ul{S}_i$) with nearest neighbor exchange constant
$J_{\mathrm{FM}}$. We introduce an easy axis in the FM ($z$-axis,
anisotropy constant $d_z = 0.02J_{\mathrm{FM}}$).  The dipolar
interaction is approximated by an additional anisotropy term
(anisotropy constant $d_x = - 0.2J_{\mathrm{FM}}$) which includes the
shape anisotropy, leading to a magnetization which is preferentially
in the $y-z$-plane.  The AFM is modeled as a magnetically diluted
Ising system with an easy axis parallel to that of the FM (Ising spin
variable $\sigma_i=\pm1$, $\epsilon_i=0,1$ depending on whether site
$i$ carries a magnetic moment or not).  Thus the Hamiltonian of our
system is given by 
\begin{align*}
{\cal H} & = - J_{\mathrm{AFM}} \!\!\!\!\sum\limits_{\langle i, j
    \rangle \in  \mathrm{AFM}}\!\!\!\! \epsilon_i \epsilon_j \sigma_i
    \sigma_j -  \sum\limits_{i \in \mathrm{AFM}} B_z \epsilon_i
    \sigma_i\\
 & -J_{\mathrm{FM}}
    \!\!\!\sum\limits_{\langle i, j \rangle \in \mathrm{FM}}
    \!\!\!{\ul S}_i \cdot {\ul S}_j - \sum\limits_{i \in \mathrm{FM}}
    \left( d_z S_{iz}^2 + d_x    S_{ix}^2 + {\ul B} \cdot {\ul S}_i
    \right)\\ 
 & -J_{\mathrm{INT}} \!\!\!\!\!\!\!\!\sum\limits_{\left\langle
    \substack{i \in \mathrm{AFM}, j \in
    \mathrm{FM}}\right\rangle}\!\!\!\!\!\!\!\!\epsilon_i \sigma_i
   S_{jz}.
\end{align*}

The first line describes the diluted AFM, the next line contains the
energy contribution of the FM and the last line includes the exchange
coupling across the interface between FM and AFM, where it is
assumed that the Ising spins in the topmost layer of the AFM interact
with the $z$ component of the Heisenberg spins of the FM. For the
nearest-neighbor exchange constant $J_{\mathrm{AFM}}$ of the AFM which
mainly determines its N\'eel temperature we set $J_{\mathrm{AFM}} = -
J_{\mathrm{FM}}/2$ and we assume $J_{\mathrm{INT}} =
-J_{\mathrm{AFM}}$. We use Monte Carlo methods with a heat-bath
algorithm and single-spin flip methods for the simulation of the model
above. The trial step of the spin update is a small variation around
the initial spin for the Heisenberg model and -- as usual -- a spin
flip for the Ising model \cite{nowakARCP01}. We perform typically
136000 Monte Carlo steps per spin for a complete hysteresis loop, a
number which turns out to be sufficient for the observed reversal
modes to be in quasi-equilibrium.  The lateral extensions of our
system are $L_y=L_z=128$ and periodical boundary conditions are used
within the film plane. The external field $\ul{B}$ is within the film
plane where $\theta$ denotes its angle to the easy axis of the FM and
AFM.

With the FM initially magnetized along the $z$ axis we first cool the
system in an external field $B = J_{\mathrm{FM}}$ with
$\theta=0^\circ$ from $\kB T = J_{\mathrm{FM}}$ to
$0.1J_{\mathrm{FM}}$ which is from above to below the ordering
temperature of the AFM. Then, for each angle $\theta$ we reset the FM
spins to be aligned along the direction of the external applied field
$\ul{B}$ and let the system relax. The initial conditions for
different angles are identical, i.\ e., we use the same defect
realization and cooling field orientation for each series of angles
and rotate the applied field $\ul{B}$ after the initial field cooling
process. The simulation of each hysteresis loop starts with a field $B
= 0.4J_{\mathrm{FM}}$ and is reduced to $-0.4J_{\mathrm{FM}}$
(decreasing branch) in steps of $\Delta B=0.004J_{\mathrm{FM}}$ before
being raised again to its initial value (increasing branch).
Hysteresis loops are simulated for angles between $0^\circ$ and
$72^\circ$ with an increment of $4^\circ$ each time taking a
configurational average over six different defect realizations.

\begin{figure}[h]
  \begin{center}
    \includegraphics[width=6.3cm]{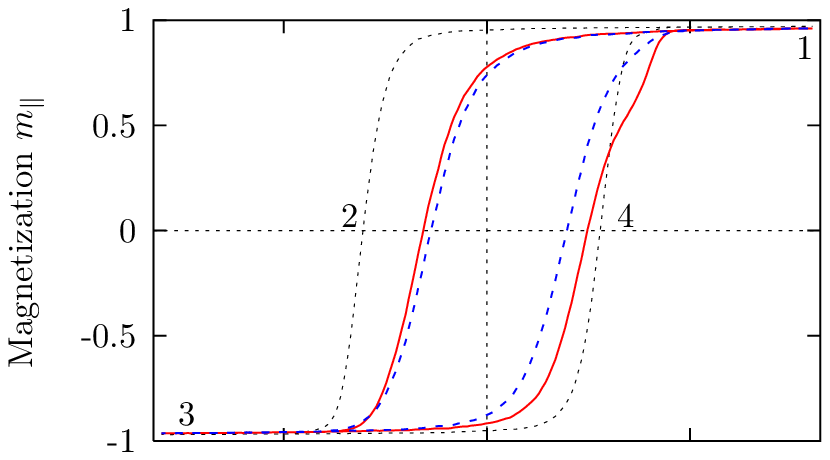}
    \includegraphics[width=6.3cm]{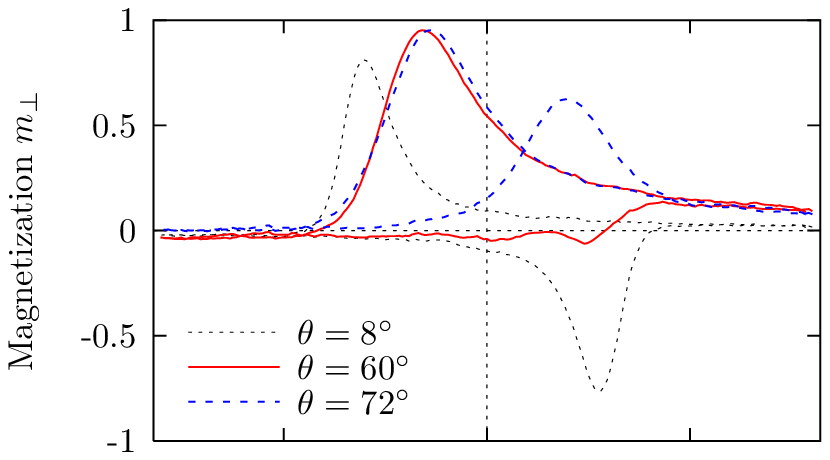}
    \includegraphics[width=6.3cm]{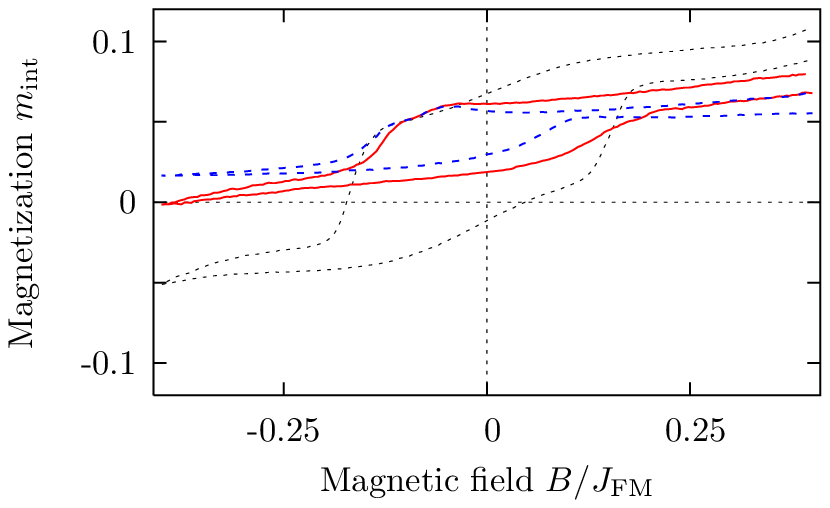}
    \end{center}
  \caption{Projection  of the magnetization along (upper graph) and
    perpendicular (central graph) to the applied field $\ul{B}$. The
  lower graph shows the AFM interface magnetization (along $z$
  direction).}
 \label{hysterese}
\end{figure}

Typical hysteresis loops for three different angles are depicted in
Fig.\ \ref{hysterese}. Shown is the projection $m_{\|}$ of the
magnetization along the applied field $\ul{B}$ (upper graph) as well
as the perpendicular component $m_{\perp}$ (center). The lower graph
shows the magnetization of the AFM interface layer (along $z$
direction). From the perpendicular component one can already see the
strong influence of $\theta$ on the reversal behavior. For small
angles we get an EB that compares to the ones of previous
simulations. With increasing angles however it becomes smaller and
even turns positive.  An explanation for this effect will be given
later. First we will focus on the reversal modes.

Depending on the angle $\theta$ one observes either a symmetric
reversal by coherent rotation ($8^\circ$), an asymmetric reversal by
coherent rotation for the descending branch and a non-uniform mode
with zero perpendicular magnetization for the ascending branch
($60^\circ$), or symmetric reversal by coherent rotation ($72^\circ$)
but -- in contrast to the usual behavior -- with the reversal along
the same direction for the descending and ascending branch.

\begin{figure}[tbh]
  \begin{center}
    \includegraphics*[bb = 110 545 290 660, width=5.2cm]{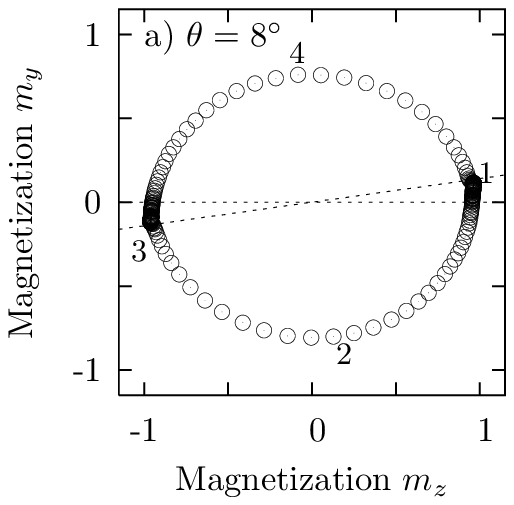}
    \includegraphics*[bb = 110 545 290 657, width=5.2cm]{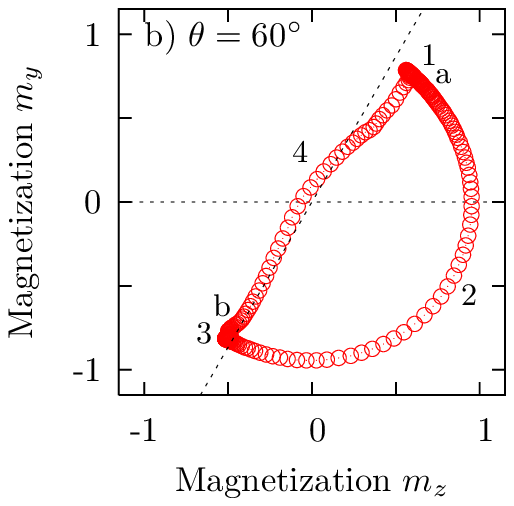}
    \includegraphics*[bb = 110 515 290 657, width=5.2cm]{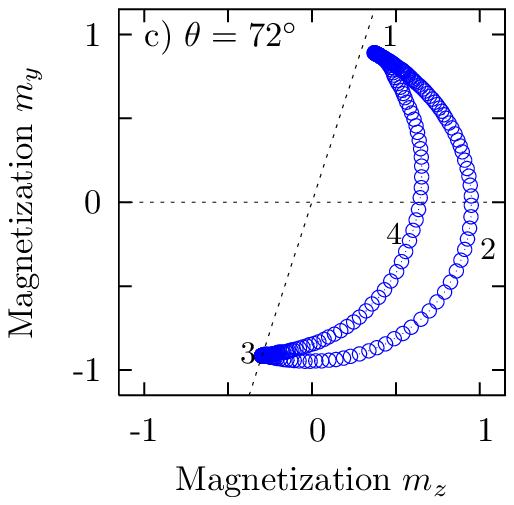}
  \end{center}
  \caption{Magnetization $m_y$ versus $m_z$ for different angles
    $\theta$; 1 and 3 indicate points with maximum external field
    $|\ul{B}|$ and 2 and 4 points where $m_{\|}$ vanishes. The
    horizontal dashed line corresponds to the easy axis of FM and AFM
    and the other dashed line to the field axis.}
  \label{circles}
\end{figure}

These effects are even more clearly shown in Fig.\ \ref{circles} where
the same angular dependence of the magnetization reversal is plotted
as $m_z$ versus $m_y$, with $m_z$ ($m_y$) denoting the magnetization
parallel (perpendicular) to the Ising axis of AFM and FM. Each plot
starts at point 1 corresponding to a magnetization with the maximum
applied external field $\ul{B}$.  Here the spins of the FM are aligned
with the external field. With decreasing field $\ul{B}$ the
magnetization starts to rotate clockwise and the parallel
magnetization will eventually vanish thus yielding the coercive field
$B^-$ (point 2). For an increasing field the corresponding field value
is $B^+$ (point 4).  When a negative maximum field is applied (point
3) the spins are aligned antiparallel with respect to their initial
orientation at point 1.

For $\theta=8^\circ$ (Fig.\ \ref{circles}a) we observe mainly
coherent rotation for magnetization reversal on both sides of the
hysteresis loop. When the external field is either decreased or
increased the FM spins first rotate towards their closest easy axis
before reversal eventually occurs.  Therefore one obtains different
signs for the perpendicular magnetization for decreasing and
increasing fields, respectively.  In Fig.\ \ref{circles}b
($\theta=60^\circ$) one observes that for a decreasing external field
the system behaves as before displaying coherent rotation. When the
external field is increased, the system shows an entirely different
behavior, though. The contribution of the magnetization perpendicular
to the field axis is negligible.  Obviously, here we have a highly
non-uniform reversal mode with vanishing total magnetization.
Increasing $\theta$ even further to $72^\circ$ (Fig.\ \ref{circles}c)
reveals again reversal mainly by coherent rotation on either side of
the hysteresis loop with a smaller net contribution $m_{\perp}$ on the
increasing branch. Most notably however is the fact that now the
rotation occurs via the same side for both branches of the hysteresis
loop, in contrast to the reversal for small angles.

\begin{figure}[tbh]
  \begin{center}
    \includegraphics[width = 4.2cm]{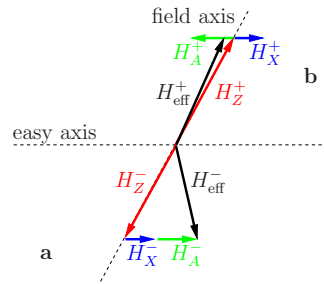}
  \end{center}
  \caption{Sketch of the effective field contributions from external
    field, $H_Z$, exchange field of AFM $H_X$, and uniaxial
    anisotropy $H_A$ acting on the FM close to the coercive fields
    $B^+$ and $B^-$ before the reversal sets in.}
  \label{asy-skizze}
\end{figure}

For an understanding of these effects we note, that there are three
different contributions to the mean effective field $\ul
H_{\mathrm{eff}}$ acting on the FM during its reversal, namely the
exchange field of the AFM, $\ul H_X = J_{\mathrm{INT}}
m_{\mathrm{AFM}} \ul{\hat{z}}$ aligned with the easy axis of the AFM,
the external magnetic field, $\ul H_Z = \ul B$, and the anisotropy
field $\ul H_A = 2 D_z m_{\mathrm{FM}} \ul{\hat{z}}$ aligned with the
easy axis of the FM, here, also the $z$ axis.

For the particular case shown in Fig.\ \ref{circles}b ($60^\circ$)
these effective field contributions are sketched in Fig.\
\ref{asy-skizze} for two characteristic values of the magnetization, a 
and b in Fig. \ref{circles}b. As is shown in Fig.\ \ref{hysterese}c),
for this value of $\theta$ the magnetization curve of the AFM which is
shifted upwards due to the fact that the AFM is in a DS with
a surplus magnetization after field cooling \cite{nowakPRB02} leads
always to a positive effective exchange field $\ul H_X$.  Additionally
an effective field coming from the anisotropy of the FM acts on the FM
which tries to rotate the magnetization to its closest easy-axis
direction. It depends on the sign of the FM magnetization and,
consequently, points into different direction on both sides of the
hysteresis. The fields labeled with $-$ in Fig.  \ref{asy-skizze}
correspond to point a in Fig. \ref{circles}b, the fields with $+$ to
point b. In the first case the effective field has a large angle with
the magnetization leading to a strong torque which favors coherent
rotation. In the second case the effective field is
more aligned with the magnetization which favors non-uniform reversal
modes. Note that this is in particular true for $\theta = 0^\circ$
where the effective field is aligned with the applied field on both
sides of the loop. Here, in agreement with the above discussion we
observed in the simulations non-uniform reversal modes on both sides.
However, we do not show these results here, since we believe that the
perfect alignment of field and easy axis is a very special case which
hardly will occur in an experimental situation.

The strength of the anisotropy field depends on the projection of
$m_{\mathrm{FM}}$ onto the easy axis and decreases with increasing
angle $\theta$. When $H_A^+$ is smaller than $H_X^+$ the reversal on
the ascending loop branch will be over the same side as that of the
descending branch (as in Fig.\ \ref{circles}c).

\begin{figure}[tbh]
  \begin{center}
    \includegraphics*[bb = 120 515 365 660, width = 6.4cm]{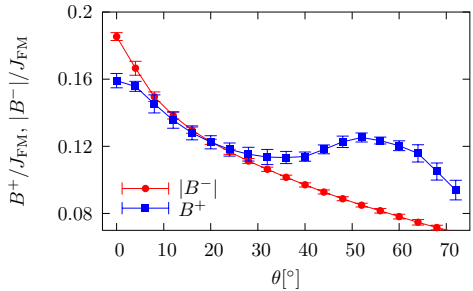}
  \end{center}
  \caption{Coercive field values $B^+$ and $B^-$ versus angle between
    field and easy axis of FM and AFM.}
  \label{hpm}
\end{figure}

A finite angle $\theta$ between the applied magnetic field and the
easy axis of the AFM can lead to another surprising effect: the
possibility of a change of the sign of the bias field when changing
this angle. The reason is a different dependence of the coercive
fields on $\theta$ due to the different reversal mechanism discussed
above. The coercive fields as function of $\theta$ are shown in Fig.\ 
\ref{hpm}. For $B^-$ the reversal mechanism is always a coherent
rotation and its value decreases with $\theta$ monotonically.  On the
other hand, for $B^+$ the behavior is more difficult. Here we found a
non-uniform reversal mode for intermediate angles for which $B^+$
increases with $\theta$ which may even lead to change of sign of the
bias field for certain angles. Note, however, that this requires bias
fields which are rather small as compared to the coercive fields, so
that this effect might not always be observed.

To summarize, varying the angle between the applied magnetic field and
the easy axis of FM and AFM we found either identical or different
reversal modes on the ascending and descending branch of the
hysteresis loop. For $\theta=60^\circ$ the reversal is maximal
asymmetric, with a coherent rotation for the descending branch and a
non-uniform reversal with zero perpendicular magnetization for the
ascending branch. We note, that different angular dependencies are
obtained for different system parameters, e.\ g.\ varying the exchange
constants or the AFM layer thicknesses.  For a comparison with
experimental situations one should note that in realistic systems the
situation might be even more complicated due to i) a twinned or
granular structure of the AFM leading to a less well defined
anisotropy axis ii) a possible additional angle between easy axis of
FM and AFM which in our investigation are identical. However, our
investigations show that in general one has to deal with different
reversal modes for the ascending and descending branch, respectively,
resulting in interesting and even surprising new effects. Furthermore,
our model reveals that a twinned crystal structure of the AFM is not
necessary for the occurrence of an asymmetry in contrast to what was
conjectured earlier \cite{fitzsimmonsPRB02}.  In corresponding
experimental investigations a systematic variation of the angle
between the magnetic field and the easy axis of FM and AFM is highly
desirable. We expect a deeper insight into the physics of EB from a
comparison of those results and our simulations.

\section*{Acknowledgments}
The authors thank B.\ Beschoten and G.\ G\"{u}ntherodt for intense
discussions.  This work has been supported by the Deutsche
Forschungsgemeinschaft through SFB 491.

\end{document}